\documentclass[aps,twocolumn,pra,showpacs,floatfix]{revtex4}
\usepackage{bm,amsmath,epsfig,graphicx,amsfonts,amssymb}
\usepackage{multirow}
\usepackage[T1]{fontenc}
\usepackage{epsfig}
\usepackage{graphicx}
\usepackage{dcolumn}
\usepackage{color}

\begin{document}
\title{Stark shift and parity non-conservation for near-degenerate states of xenon}

\author{L. Bougas}
\affiliation{Department of Physics, University of Crete - 71003 Heraklion-Crete, Greece }%
\affiliation{Institute of Electronic Structure and Laser, Foundation for Research and Technology-Hellas 
71110 Heraklion-Crete, Greece}%

\author{G. E. Katsoprinakis}
\affiliation{Department of Physics, University of Crete - 71003 Heraklion-Crete, Greece }%
\affiliation{Institute of Electronic Structure and Laser, Foundation for Research and Technology-Hellas 
71110 Heraklion-Crete, Greece}%

\author{D. Sofikitis}
\affiliation{Department of Physics, University of Crete - 71003 Heraklion-Crete, Greece }%
\affiliation{Institute of Electronic Structure and Laser, Foundation for Research and Technology-Hellas
 71110 Heraklion-Crete, Greece}%

\author{P. C. Samartzis}
\affiliation{Department of Chemistry, University of Crete - 71003 Heraklion-Crete, Greece }%
\affiliation{Institute of Electronic Structure and Laser, Foundation for Research and Technology-Hellas
 71110 Heraklion-Crete, Greece}%
\author{T. N. Kitsopoulos}
\affiliation{Department of Chemistry, University of Crete - 71003 Heraklion-Crete, Greece }%
\affiliation{Institute of Electronic Structure and Laser, Foundation for Research and Technology-Hellas 
71110 Heraklion-Crete, Greece}%

\author{T. P. Rakitzis}
\email{ptr@iesl.forth.gr}
\affiliation{Department of Physics, University of Crete - 71003 Heraklion-Crete, Greece }%
\affiliation{Institute of Electronic Structure and Laser, Foundation for Research and Technology-Hellas
 71110 Heraklion-Crete, Greece}%

\author{J. Sapirstein}
\address{Department of Physics, University of Notre Dame, Notre Dame, Indiana 46556 }%

\author{D. Budker}
\affiliation{Helmholtz Institute, Johannes Gutenberg University, 55099 Mainz, Germany}
\affiliation{Department of Physics, University of California, Berkeley California 94720-7300, USA}%
\affiliation{Nuclear Science Division, Lawrence Berkeley National Laboratory, Berkeley California 94720, USA}%

\author{V. A. Dzuba}
\affiliation{School of Physics, University of New South Wales,
Sydney, NSW 2052, Australia}
\author{V. V. Flambaum}
\affiliation{School of Physics, University of New South Wales,
Sydney, NSW 2052, Australia}
\author{M. G. Kozlov}
\affiliation{School of Physics, University of New South Wales,
Sydney, NSW 2052, Australia}

\date{ \today}


\begin{abstract}
We identify a pair of near-degenerate states of opposite parity in atomic Xe, the $5p^5 10s \,\, ^2[3/2]_2^o$ at $\rm{E}=94759.927$\,cm$^{-1}$ and $5p^5  6f \,\, ^2[5/2]_2$ at $\rm{E}= 94759.935$\,cm$^{-1}$, for which parity- and time-odd effects are expected to be enhanced by the small energy separation. We present theoretical calculations which indicate narrow widths for both states and we report a calculated value for the weak matrix element, arising from configuration mixing, of $|W|=2.1$\,Hz for $^{132}$Xe. In addition, we measured the Stark effect of the $5p^5\,6f$ $^2[5/2]_{2}$ and $5p^5 \,6f \ ^2[3/2]_2$ ($\rm{E} =94737.121\,\rm{cm}^{-1}$) states. The Stark-shift of the $6f$ states is observed to be negative, revealing the presence of nearby $6g$ states at higher energies, which have not been observed before. The Stark-shift measurements imply an upper limit on the weak matrix element of $|W|\!<\!5$\,Hz for the near-degenerate states ($10s \,\, ^2[3/2]_2^o$ and $6f \,\, ^2[5/2]_2$), which is in agreement with the presented calculations.

\end{abstract}

\maketitle
\section{Introduction}
\begin{figure}[h!]
\begin{center}
$\begin{array}{l@{\hspace{0.21in}}r}
\multicolumn{1}{l}{\mbox{(a)}} &\\
\includegraphics[angle=0, width=0.9\linewidth]{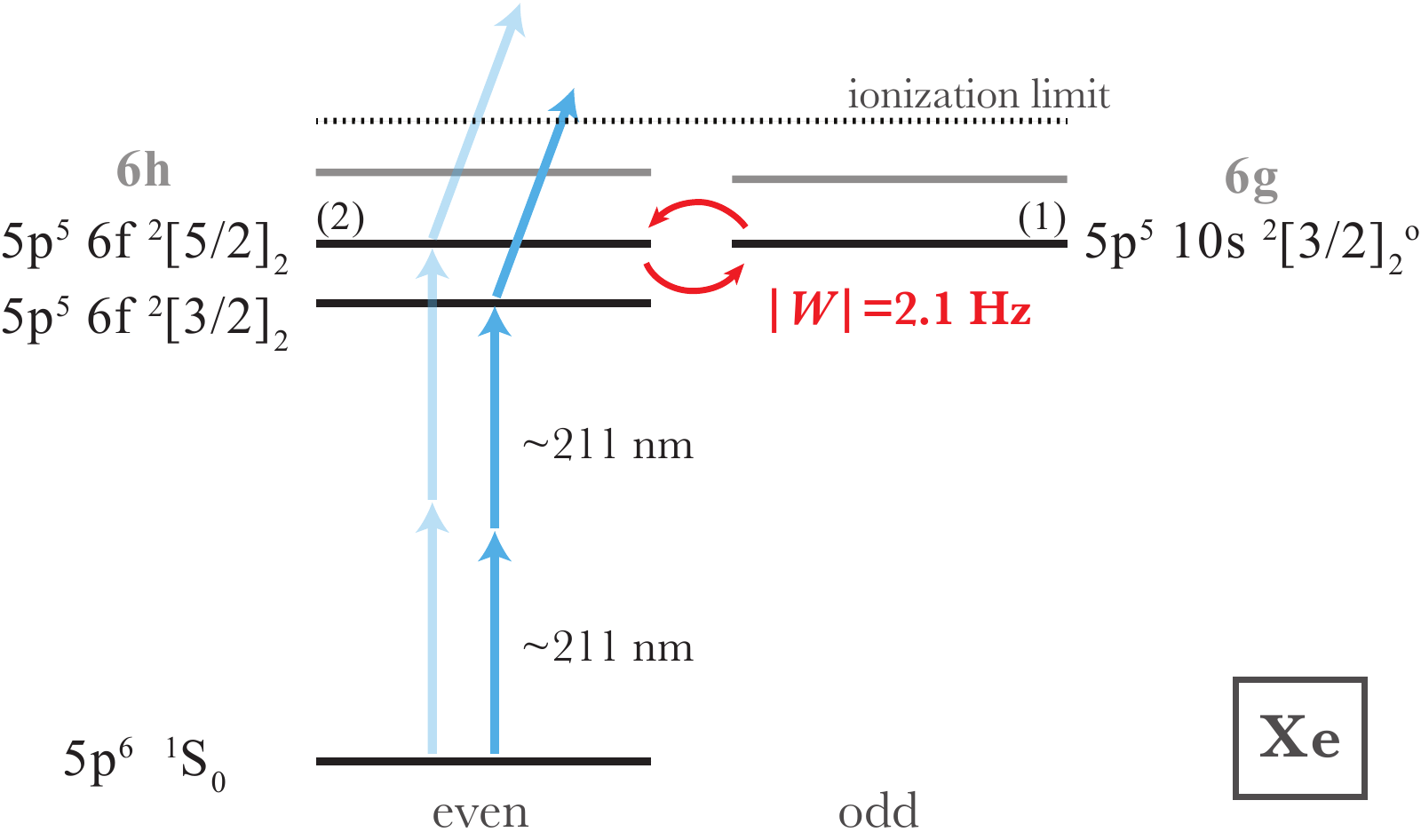}&\\[+0.2cm]
	\multicolumn{1}{l}{\mbox{(b)}} &\\ 
	\includegraphics[angle=0, width=0.9\linewidth]{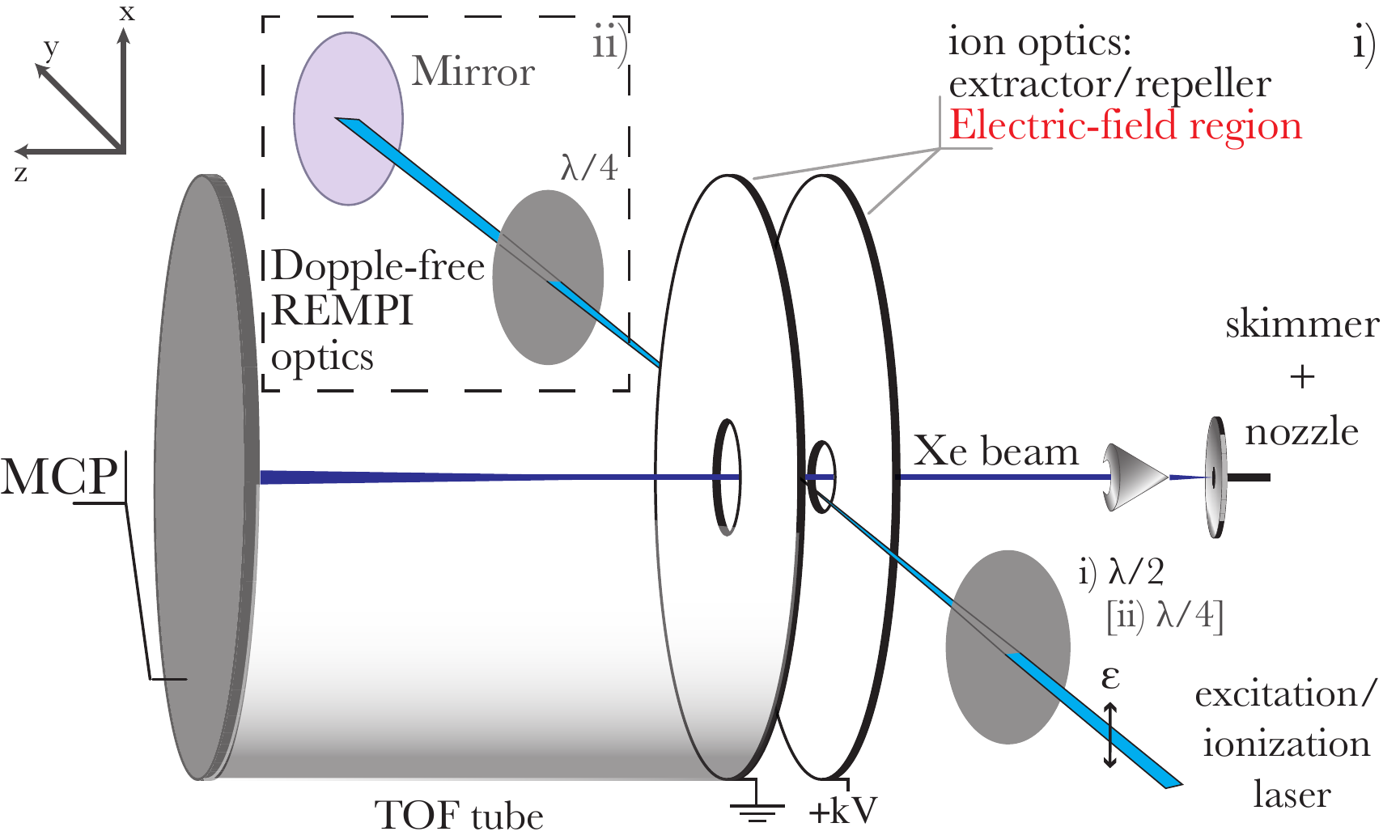}&\\[-0.2cm]
\end{array}$
\end{center}
\caption{\small{(color online) a) Partial energy-level diagram of Xe (not to scale) showing the pair of near-degenerate states \ref{odd} and \ref{even}. The solid lines represent the $\sim$\,211\,nm radiation used for the measurements of the Stark effect of the $6f$ states using a (2+1) REMPI scheme, see Sec. \ref{sec:Results}. The analysis of the measurements of the Stark effect of the two 6$f$ states, reveals the presence of the $6g$ and $6h$ states, which have not been observed before. b) Schematic drawing of the experimental setup used for the Stark-shift measurements. i) A collimated Xe beam is intersected at right angles by a linearly polarized light beam from a $\sim$\,211\,nm pulsed laser. The laser pulse excites the atoms from the ground state to the excited state of interest, and subsequently ionizes the atoms. The electric field along the $z$-axis, is used to accelerate the ions down the time-of-flight (TOF) tube towards a microchannel-plate (MCP) detector; this is also the electric field in the Stark effect measurements. The polarization of the laser beam is controlled with a $\lambda/2$ plate. For excitation to the $|M|=2$ sublevels (with quantization axis along $\hat{z}$) of the $6f$ excited states, the polarization of the laser beam is set perpendicular to the $z$-axis. ii) Additional optics used only for the calibration of the electric field, which is realized through the Stark-shift measurements of the $n=2$ states in H using a Doppler-free REMPI scheme (see discussion in Sec.\,\ref{seq:EfieldCalib}).}}
\label{fig:ExpApparatus}
\end{figure}

\indent High-Z atoms with near-degenerate opposite-parity states are promising spectroscopic systems for experiments studying parity (P) and time-reversal invariance (T) violating phenomena, due to the enhancement of the P- and P,T-odd effects by the small energy intervals between the atomic levels\,\cite{Ginges}. In 1986, Dzuba, Flambaum and Khriplovich\,\cite{Dzuba1986} identified the importance of various pairs of near degenerate states of opposite parity in rare earth atoms, such as samarium (Sm), erbium (Er) and dysprosium (Dy), for possible enhancement of such effects. In particular, atomic Dy has been used in measurements of parity nonconservation (PNC)\,\cite{Nguyen} and also in search for the time variation of the fine-structure constant\,\cite{Cingoz}, and to investigate the gravitational-potential dependence on the fine-structure constant\,\cite{Ferrell}, with more recent results additionally setting limits on violations of Lorentz symmetry and the Einstein equivalence principle\,\cite{BudkerLE}. \\
\indent Here, we identify a pair of near-degenerate opposite-parity states, both with total angular momentum $J=2$, in atomic xenon (Xe) as a candidate system for P- and T-violation experiments, Fig.\,\ref{fig:ExpApparatus}\,(a); see Ref.\,\cite{NIST}:
\begin{align}
5p^5 \,10s\,\, & ^2[3/2]_2^o & {\rm E=} 94759.927 \ {\rm cm}^{-1}, \label{odd}\\
5p^5  \,6f \,\,& ^2[5/2]_2   & {\rm E=} 94759.935 \ {\rm cm}^{-1}. \label{even}
\end{align}
\indent Xe is a noble gas with nuclear charge Z=54 and therefore the $Z^3$ enhancement of P- and T-odd effects is significant for the chosen states\,\cite{BouchiatBouchiat}. The wide range of stable Xe isotopes  (eight, ranging from $^{129}$Xe to $^{136}$Xe) gives the possibility to perform PNC measurements in different isotopes, in which the large error associated with atomic theory calculations can be largely eliminated by taking ratios of measurements in different isotopes\,\cite{Ginges,Dzuba1986}. Additionally, two of the stable Xe isotopes have nonzero nuclear spin ($^{129}$Xe with $I=1/2$ and $^{131}$Xe with $I=3/2$), and therefore PNC measurements in different hyperfine components give access to the nuclear anapole moment\,\cite{Khriplovich}. In close analogy to experiments in Dy, possible searches for variations of the fine-structure constant using this pair of near degenerate states, may become feasible due to the wide choice of rf transitions, arising from the availability of many isotopes. \\
\indent The principal motivation for this work is the prospect of measuring parity violating phenomena using the near-degenerate opposite-parity states \eqref{odd} and \eqref{even}. Parity violation in atomic systems arises primarily due to the weak interaction between the electron and the nucleus, which results in mixing of the parity-eigenstates of the system. In particular, the weak interaction mixes predominantly $s$ and $p$ states\,\cite{BouchiatBouchiat,Khriplovich}. Therefore, mixing arising from PNC between the dominant configurations ($6f-10s$) of the levels \eqref{odd} and \eqref{even} is negligible. However, if the 6$f$ state has significant $p$ character from configuration mixing with nearby $np$ states (dominated by the nearest $10p$), then the weak interaction will mix the opposite-parity states \eqref{odd} and \eqref{even}. In this article, we present theoretical calculations of the composition of the states \eqref{odd} and \eqref{even}, from which we can estimate the electric-dipole matrix element between these states, as well as the matrix element arising from the weak interaction, both required to assess the possibility of performing PNC experiments using the Stark-interference technique\,\cite{Nguyen}. In addition, experimental measurements of the Stark-shift of the $6f$ state \eqref{even} are presented, which support the accuracy of the presented theoretical calculations.\\
\indent In particular, in Sec.\,\ref{sec:Theory} we present calculations of the width of the states \eqref{odd} and \eqref{even}, which are important for assessing the experimental sensitivity for future P- and T-odd experiments, and of the matrix elements of the electric-dipole interaction and of the weak interaction between these states. In addition, we study the magnetic dipole transitions involving our states of interest, as these are sensitive to the composition of states \eqref{odd} and \eqref{even}. In Sec.\,\ref{sec:Stark} we present measurements of the Stark effect of two $6f$ states (at $94737.121$\,cm$^{-1}$ and at $94759.935$\,cm$^{-1}$) with the aim to determine the admixture of the  $5p^5\,np$ configurations to the $5p^5\,6f$ configuration. Finally, the analysis of the Stark measurements, performed in Sec.\,\ref{sec:Results}, determines an upper limit on the weak mixing between states \eqref{odd} and \eqref{even}, in agreement with the presented theoretical calculations.\\

\section{Theory}
\label{sec:Theory}
\subsection{Width of the states}
\indent The energy difference between the states (\ref{odd}) and (\ref{even}) is
only 0.008 cm$^{-1}$. Such a small value might be comparable to the
linewidth. The requirement that the widths of the states involved are small, is of importance for
PNC measurements.
Therefore, we estimate the widths of both
states. The natural width is determined by the electric dipole
transitions from the states (\ref{odd}) or (\ref{even}) into appropriate
lower states (we use atomic units: $|e| = \hbar = m = 1$)
\begin{equation}
\Gamma_a = \sum_b \frac{4}{3} \alpha^3 \omega_{ab}^3 \frac{A_{ba}^2}{2J_a+1},
\label{eq:game1}
\end{equation}
where $a$ stands for (\ref{odd}) or (\ref{even}), while $b$ represents all possible lower states connected to (\ref{odd}) or (\ref{even}) via the electric dipole operator, $\omega_{ab}$ is the transition frequency, $J_a$ is the total angular momentum quantum number of state $a$, and $A_{ba}$ is the transition probability from state $a$ to $b$.\\
\indent For estimations we consider the $10s$ - $6p$ and $6f$ - $5d$
transitions in the single-electron approximation. Transitions to higher
states ($7p,8p,6d$, etc.) can be neglected because of the smaller
frequencies of the transitions. For example, the suppression of the $10s -7p$
transition probability compared to the $10s -6p$ transition probability is approximately a factor of
 $\sim20$, i.e. $\left[(E(10s)-E(6p))/(E(10s)-E(7p)\right]^3 \sim
20$. This cannot be compensated by the larger transition amplitude since the
corresponding ratio, as calculations show, is much smaller, $[A(10s-7p)/A(10s-6p)]^2 \sim 3$.\\
\indent Taking the electric dipole
amplitudes from the random-phase approximation (RPA) calculations (see, for example Ref.\,\cite{CPM}): $A_{10s,6p} = 0.5 \ ea_0$,
$A_{6f,5d} = 3 \ ea_0$, and assuming $\omega_{10s,6p} \approx 17000$
cm$^{-1}$ = 0.08 a.u., $\omega_{6f,5d} \approx 15000$ cm$^{-1}$ =
0.07 a.u leads to $\Gamma_{10s} = 1.3 \times 10^{-11}$ a.u. = $3
\times 10^{-6}$ cm$^{-1}$, and $\Gamma_{6f} = 3 \times 10^{-10}$ a.u. = $6
\times 10^{-5}$ cm$^{-1}$. \\
\indent To get a more precise estimate, we turn to the configuration interaction
(CI) method. We calculate the transition amplitudes from the $5p^5\,10s$\,$^2[3/2]_{J=2}^o$ state to five lower states of the $5p^5\,6p$ configuration and
from the $5p^5\,6f$ $^2[5/2]_{J=2}$ state to six states of
the $5p^55d$ configuration. The calculated E1 transition amplitudes are
presented in Table~\ref{t:ampE1}. The main source of uncertainties is
the incompleteness of the basis. Given that the main contributions are
included and that the results are stable against variation of the basis,
we estimate that the uncertainty for the E1 transition amplitudes should not exceed
20\%. The uncertainties for the linewidths are larger because the
amplitudes are squared in the transition probabilities and because higher lying states, to which decay can occur, are neglected in the calculations. We estimate the uncertainties for the linewidth calculations to be $\sim$ 50\%.\\
\indent The results for the linewidths of states (\ref{odd}) and (\ref{even}) are listed in Table\,I. The corresponding lifetimes are $\tau_{10s} = 500(250)$ ns,
$\tau_{6f} = 170(85)$ ns. The theoretical result for the lifetime of the 6$f$ state is consistent with the experimental value of 153(12) presented in Ref. \cite{Verdugo}. \\

\begin{table}
\caption{Electric dipole transition amplitudes (reduced matrix
  elements, $A_{ab}$ in atomic units) from the
  degenerate states (\ref{odd}) and (\ref{even}) to lower states and the
  corresponding linewidths $\Gamma_a$.}
\label{t:ampE1}
\begin{ruledtabular}
\begin{tabular}{lcclccl}
From ($a$) & \multicolumn{3}{c}{$5p^510s, J=2$} &
       \multicolumn{3}{c}{$5p^56f, J=2$} \\
to ($b$)  & \multicolumn{3}{c}{$5p^56p$} &
       \multicolumn{3}{c}{$5p^55d$} \\
& $J_b$ & $E_b$(cm$^{-1}$) & \multicolumn{1}{c}{$A_{ab}$} &
$J_b$ & $E_b$(cm$^{-1}$) & \multicolumn{1}{c}{$A_{ab}$} \\
\hline
& 1 & 77269 & 0.44(9)   & 1 & 79987 & 1.1(2) \\
& 2 & 78119 & 0.32(6)   & 3 & 80970 & 0.35(7) \\
& 3 & 78403 & 0.72(14)  & 2 & 80323 & 0.65(13) \\
& 1 & 78956 & 0.14(3)   & 1 & 83890 & 2.0(4) \\
& 2 & 79212 & 0.519(10)  & 2 & 81925 & 1.3(3) \\
&   &       &       & 3 & 82430 & 0.24(5) \\
$\Gamma_a$ & & $\sim 1\times 10^{-5}$ &&& $ \sim 3 \times 10^{-5}$ & \\
\end{tabular}
\end{ruledtabular}
\end{table}

\subsection{PNC and Stark mixing} 
\indent One possibility in using states (\ref{odd}) and (\ref{even}) for
PNC studies in xenon is to perform an experiment similar to that in dysprosium~\cite{Nguyen}, in which the interference between Stark mixing and PNC mixing of the two degenerate states of opposite
parity was studied. The mixing is determined by the matrix elements
of the weak interaction and the electric-dipole interaction between states
(\ref{odd}) and (\ref{even}),
\begin{equation}
W = \langle 1 |H_W| 2 \rangle \ \ {\rm and} \ \ D = -\langle 1 |e{\mathbf
  r}| 2 \rangle.  
\label{WD}
\end{equation}
The Hamiltonian $H_W$ of the weak interaction is given by
\begin{equation}
  H_W = -\frac{G_F}{2\sqrt{2}} Q_W \gamma_5 \rho(r).
\label{eq:hw}
\end{equation}
The $G_F$ in (\ref{eq:hw}) is the Fermi constant of
the weak interaction ($G_F \approx 2.2225 \times 10^{-14}$ a.u.),
$Q_W$ is the nuclear weak charge,
$\gamma_5$ is a Dirac matrix, and $\rho({\bf r})$ is the
nuclear density normalized to 1 ($\int\rho dV=1$).

\indent Within the standard model
the weak nuclear charge $Q_W$ is given by~\cite{SM}
\begin{equation}
Q_W \approx -0.9877N + 0.0716Z.
\end{equation}
Here $N$ is the number of neutrons, $Z$ is the number of protons.
For example, for $^{132}$Xe, $Q_W = -73.17$.

\begin{figure}[h!]
\centering
\epsfxsize=3in
\epsffile{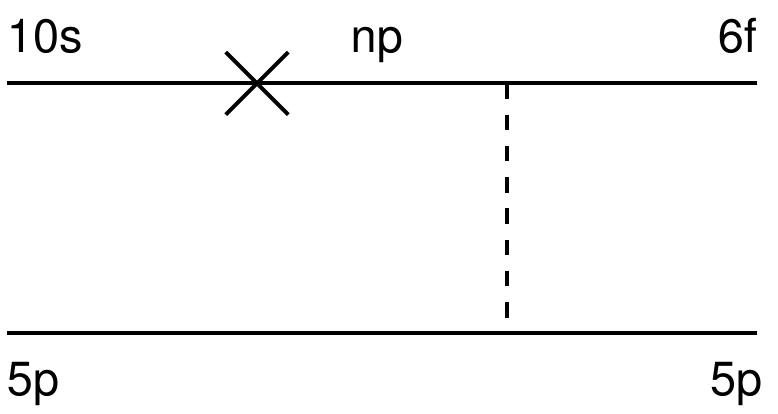}
\caption{\small{The dominant diagram for the weak matrix element between the near-degenerate
states (\ref{odd}) and (\ref{even}). Dashed line is the Coulomb
interaction, and the cross stands for the weak interaction. Other diagrams can be
obtained by adding an exchange diagram (swapping states $6f$ and
$5p$) and by moving the operator of the weak interaction in turn to every
other electron line. Similar diagrams describe the electric dipole
transition amplitude. In the latter case, the cross would stand for the
electric dipole operator.}}
\label{f:mix}
\end{figure}

\indent The matrix elements $W$ and $D$ in Eq. (\ref{WD}) have non-zero values due to
configuration mixing in either odd- and even-parity states. Indeed, 
the parity violating matrix elements of the weak interaction are
nonzero only for states of the same total angular momentum $J$, opposite
parity, and whose configurations are different only by one
electron. For the electric dipole operator the selection rule for the
total angular momentum is $|\Delta J| \leq 1$. For such 
configurations, the matrix element between many-electron states is 
reduced to the single-electron matrix element between two different
single-electron states. \\
\indent The nominal configurations of states (\ref{odd}) and (\ref{even}) differ by the $10s$ and $6f$
orbitals. The matrix elements of the weak ($W$) and the electric dipole ($D$) interactions
between these states are zero. One has to include mixing of appropriate configurations to
either of the states (\ref{odd}) or (\ref{even}) to obtain non-zero
matrix elements. The dominant contribution arises when mixing with the
$5p^5np_{1/2}$ configurations is included in the wave function of the
$5p^56f$ state (\ref{even}). The corresponding diagram is shown in
Fig.~\ref{f:mix}. There are seven similar diagrams that can be
obtained from Fig.~\ref{f:mix} by moving the operator of the weak or
the electric dipole interaction in turn to every electron line and by
swapping the $6f$ and $5p$ orbitals on the right.

\indent To calculate the matrix element corresponding to the diagram, we need
to know the many-electron wave function of each of the initial and final states (\ref{odd}) and
(\ref{even}). It is convenient to use the hole-particle representation for
the wave functions. In this representation, the wave function of the
odd-parity state (\ref{odd}) is simple. It contains one $10s$ electron and
one $5p_{3/2}$ hole. Since the total angular momentum is two, there is
only one way to obtain $J=2$ and $J_z=2$ - by combining the
$10s_{1/2,1/2}$ electron state with the $5p_{3/2,3/2}$ hole
state. The corresponding wave function is just the product of two
single-particle wave functions.

\indent The $5p6f, \ J=2$ state is more
complicated and standard configuration interaction (CI) techniques and
angular momentum algebra are used to construct it. The expansion of this state
over states composed of the single-electron state $6f$ and the single-hole
state $5p$ is presented in Table~\ref{t:6f}.

\begin{table}
\caption{Expansion of the state $5p6f \ J=2,J_z=2$ over states with different
  total angular momentum $j$ and its projection $j_z$ ($|5p6f
  \ ^2[5/2]_{22} \rangle = \sum_i c_i |6f_{j_im_i}5p_{j^{\prime}_im^{\prime}_i}\rangle$).}
\label{t:6f}
\begin{ruledtabular}
\begin{tabular}{crcccr}
 & & \multicolumn{2}{c}{$6f$} &  \multicolumn{2}{c}{$5p$} \\
 $i$ & \multicolumn{1}{c}{$c_i$} & $j_i$ & $m_i$& $j^{\prime}_i$ &
 \multicolumn{1}{c}{$m^{\prime}_i$}  \\
\hline
1 &  0.046866 & 7/2 &  1/2 & 3/2 &  3/2 \\
2 & -0.104795 & 7/2 &  3/2 & 3/2 &  1/2 \\
3 &  0.181510 & 7/2 &  5/2 & 3/2 & -1/2 \\
4 & -0.277262 & 7/2 &  7/2 & 3/2 & -3/2 \\
5 &  0.000855 & 5/2 &  3/2 & 1/2 &  1/2 \\
6 & -0.001912 & 5/2 &  5/2 & 1/2 & -1/2 \\
7 &  0.353957 & 5/2 &  1/2 & 3/2 &  3/2 \\
8 & -0.578009 & 5/2 &  3/2 & 3/2 &  1/2 \\
9 &  0.646233 & 5/2 &  5/2 & 3/2 & -1/2 \\
\end{tabular}
\end{ruledtabular}
\end{table}

\indent The mathematical expression for the diagram presented on Fig.~\ref{f:mix} is the following:
\begin{equation}
W_1 = \sum_{i,n} c_i \frac{\langle 10s| \tilde{H}_W| np \rangle
  \langle np,5p |\frac{e^2}{r_{12}}|
  6f_{j_i,m_i}5p_{j^{\prime}_i,m^{\prime}_i} \rangle}{\epsilon_{10s} -
  \epsilon_{np}}.
\label{eq:w1}
\end{equation}
Here $c_i$ are the expansion coefficients from Table~\ref{t:6f},
$\tilde{H}_W = H_W + \delta V_W$,
$H_W$ is the Hamiltonian of the weak interaction (\ref{eq:hw})
and $\delta V_W$ is the correction to the self-consistent Hartree-Fock
potential due to the weak interaction (RPA-type correction~\cite{CPM}).
The expressions for the other seven diagrams are similar to (\ref{eq:w1}).

\indent To calculate the weak matrix element $W$ ($W = \sum_{i=1}^{8} W_i$)
we use the relativistic Hartree-Fock (RHF) method and the B-spline
technique~\cite{Bspline}. The calculations are performed in the $V^{N-1}$
approximation with one electron removed from the uppermost $5p$ core
shell. The states of the external electron are calculated in the frozen
$V^{N-1}$ potential using the B-spline technique. Summation over $np$
in (\ref{eq:w1}) goes over the complete set of single-electron states
constructed with the use of B-splines.

\indent The result for $^{132}$Xe is
\begin{equation}
  |W| = 0.34 \times 10^{-15} (-Q_W/N) \ \rm{a.u.} = 2.1 \ {\rm Hz}.
\label{eq:res}
\end{equation}
Note that the dominating term $W_1$ (\ref{eq:w1}) gives 105\% of the
matrix element while other seven diagrams give only -5\%.

\indent Similar calculations for the reduced electric-dipole matrix element $D$, lead to
\begin{equation}
  A(E1) = \langle 10s5p J=2|| E1 || 6f5p J=2 \rangle = 1.55 \ ea_0,
\label{eq:e1}
\end{equation}
or
\begin{equation}
  A(E1)_z = 0.57 \ ea_0 \rightarrow 0.024\,\rm{cm}^{-1}/(\rm{kV/cm})
\label{eq:e1z}
\end{equation}
for the electric dipole transition amplitude between states (\ref{odd}) and (\ref{even}).\\

\subsection{Magnetic-dipole transition amplitudes} 
\begin{table}
\caption{Experimental and calculated $g$-factors of the states of the $5p^55d$, $5p^510s$,
  $5p^56p$, and $5p^56f$ configurations with total angular momentum
  $J=2$.}
\label{t:gf}
\begin{ruledtabular}
\begin{tabular}{rcc ll}
$N$ & Configu- & Energy &\multicolumn{2}{c}{$g$-factors} \\
& ration & \multicolumn{1}{c}{cm$^{-1}$} & \multicolumn{1}{c}{Experiment}
& \multicolumn{1}{c}{Calculations} \\ 
\hline
1 & $5p^55d$ & 80323 & 1.3750 & 1.35 \\
2 &     \text{\textquotedbl}    & 81925 &    -    & 0.95 \\
3 &      \text{\textquotedbl}    & 91153 &    -    & 0.79 \\
4 &      \text{\textquotedbl}    & 91447 & 1.274  & 1.25 \\
5 & $5p^510s$& 94760 & 1.512  & 1.50 \\
6 & $5p^56p$ & 78120 & 1.11103& 1.10 \\
7 &      \text{\textquotedbl}    & 79212 & 1.3836 & 1.39 \\
8 &    \text{\textquotedbl}      & 89162 & 1.190  & 1.17 \\
9 & $5p^56f$ & 94737 & 1.09   & 1.11 \\
10&   \text{\textquotedbl}       & 94760 & 0.87   & 0.84 \\
\end{tabular}
\end{ruledtabular}
\end{table}

\begin{table}
\caption{Calculated magnetic-dipole transition amplitudes (reduced matrix elements given in units of Bohr magneton $\mu_0$) between the states numerated in Table~\ref{t:gf} (labeled here as $N$ and $N^{\prime}$).}
\label{t:m1}
\begin{ruledtabular}
\begin{tabular}{cl cl}
$N-N^{\prime}$ & \multicolumn{1}{c}{$M1$} &
$N-N^{\prime}$ & \multicolumn{1}{c}{$M1$} \\
\hline
\multicolumn{2}{c}{$10s - 5d$ transitions} &
\multicolumn{2}{c}{$6f - 6p$ transitions} \\
5 - 1 & 0.20(4)  &  9 - 6 & 0.0064(13) \\
5 - 2 & 0.052(10) &  9 - 7 & 0.023(5) \\
5 - 3 & 0.023(5) &  9 - 8 & 0.0093(19) \\
5 - 4 & 0.26(5)  & 10 - 6 & 0.0030(6) \\
\multicolumn{2}{c}{$6f - 6f$ transition} & 10 - 7 & 0.0011(2) \\ 
9 -10 & 0.19(4)  & 10 - 8 & 0.0089(18) \\
\end{tabular}
\end{ruledtabular}
\end{table}

\indent It is instructive to study the magnetic dipole transitions ($M1$) involving
our states of interest, (\ref{odd}) and (\ref{even}). The $M1$ amplitudes are sensitive to the composition of the states of interest, similar to the $E1$ transition amplitudes and the weak matrix element $W$, and therefore provide valuable information about the composition of the states. Knowledge of these matrix elements is also needed to assess the systematic effects in PNC experiments (see for example Ref.\,\cite{Tsigutkin}). Moreover, here we have
better control over the accuracy of the calculations, since the calculation of
the $M1$ amplitudes is similar to the calculation of the
$g$-factors, and $g$-factors of many states of xenon are known from
experiment. \\
\indent In this section we study the M1 transitions between the odd-parity state (\ref{odd})
and states of the $5p^56d$ configuration, and the even-parity state (\ref{even}) and
states of the $5p^56p$ configuration. We limit ourselves to states
of total angular momentum $J=2$. The former transitions can be
described as the $10s$ - $5d$ single-electron transitions while the
later as the $6f$ - $6p$ single-electron transitions.
The corresponding single-electron matrix elements are relatively small,
\begin{eqnarray}
\langle 10s_{1/2} ||M1||5d_{3/2} \rangle = 0.0107 \mu_0, \label{m1sd} \\ 
\langle 6f_{5/2} ||M1||6p_{3/2} \rangle = 0.0067 \mu_0,
\label{m1fp} 
\end{eqnarray}
where $\mu_0$ is the Bohr magneton. These values were obtained in the RPA calculations. They are dominated
by the core-polarization effect (also known as the RPA correction).\\
\indent At first glance, the transitions between many-electron states should be
reduced to the matrix elements \eqref{m1sd} and \eqref{m1fp} with some
modification due to configuration mixing. In fact, the nature of the
many-electron transitions is completely different and the matrix elements
(\ref{m1sd}) and (\ref{m1fp}) play little role in it. For example, in the case of the $10s$ - $5d$, the single-electron transition amplitudes when configuration mixing is included can be reduced to:
\begin{eqnarray}
\langle \tilde{10s} ||M1||\tilde{5d}\rangle=&\alpha \langle 10s ||M1||5d\rangle+\beta \langle 10s ||M1||10s\rangle \nonumber\\
& + \,\gamma \langle 5d ||M1||5d\rangle.
\label{eq:M1config}
\end{eqnarray}
For pure configurations $\alpha=1$ and $\beta=\gamma=0$. When configuration mixing is included $\alpha\sim1$; $\beta$,\,$\gamma\ll1$. However, the second and third terms in Eq. (\ref{eq:M1config}) dominate over the first one in spite of the small values of $\beta$ and $\gamma$, because these $M1$ matrix elements have values of $\sim \mu_0$, as these belong to transitions between states with the same principal and angular quantum numbers $n,l$ (i.e. diagonal in $n,l$) while also satisfying $\Delta j=0,\pm1$. 
This leads us to the contributions which have diagonal
$M$1 single-electron matrix elements or matrix elements between the fine-structure partners. 
Therefore, an estimate for the $5p^510s$ - $5p^55d$
transitions can be written as:
\begin{equation}
  M1 \approx \frac{\tilde g(10s,5p,5d,5p)}{\epsilon_{10s}-\epsilon_{5d}}
  \mu_0,
\label{eq:m1}
\end{equation}
where $\tilde g(a,b,c,d) = g(a,b,c,d) - g(a,b,d,c)$ is the sum of
the direct and the exchange Coulomb integrals. Substituting these numbers leads to $M1 \sim \mu_0.\\$
\indent For a more accurate estimate we turn to the CI calculations and we
start from the $g$-factors to get some idea about the accuracy of the
calculations. The values of the $g$-factors can be expressed via
diagonal $M$1 matrix elements as:
\begin{eqnarray}
  g_a = \frac{1}{J_a} \left(\begin{array}{rrr} J_a & 1 & J_a \\ -J_a &
      0 & J_a \\ \end{array} \right) \langle a||M1||a \rangle,
\label{g-factor}
\end{eqnarray}
where $J_a$ is the total angular momentum of the state $a$.

\indent The results for the $g$-factors are presented in Table~\ref{t:gf}
together with available experimental data. The deviation of the theory
from the experiment does not exceed 4\%. Given that the accuracy for the diagonal
matrix elements is often higher than that for the off-diagonal ones, we can
say that this finding is consistent with our earlier estimate of the
20\% uncertainty for the electric dipole matrix elements. To be on the
safe side we adopt the same 20\% uncertainty for the $M$1 transition
amplitudes. The results are presented in Table\,\ref{t:m1}. Note that
some amplitudes for the transitions between states of the $5p^510s$
and $5p^55d$ configurations are larger than the single-electron matrix
element\,\eqref{m1sd}. One can say that configuration mixing
can lead to enhancement of the $M$1 transition amplitudes.

\subsection{Comparison with dysprosium}
\indent It is interesting to compare the weak matrix element between the
degenerate states of Xe [\eqref{odd} and \eqref{even}], with the weak matrix elements of the degenerate states
A and B for Dy~\cite{Nguyen,DzubaDy}. The experimental value for Dy
is $|W_{\rm AB}| =| 2.3 \pm 2.9 \pm 0.7|$ Hz~\cite{Nguyen}, while
the theoretical value is  $W_{\rm
  AB} = 4(4) \ {\rm Hz}$~\cite{DzubaDy}. Note that highly excited $np_{1/2}$
states give significant contribution to the value (\ref{eq:res}) for
the weak matrix element for Xe, making it smaller. Such states were
not included into calculation of the weak matrix element for Dy. 
At the present level of accuracy the predicted values of the PNC matrix elements for these two systems are similar.

\section{Stark Effect of the $6f$ states}
\label{sec:Stark}
\indent To study the states of prospective PNC measurements, the Stark shift of
the energies of two $6f$ states with total angular momentum
$J=2$ in a uniform dc electric field were measured. One state is the $5p^5\,6f\,\,^2[5/2]_{2}$ with energy $\rm{E}=94759.935$ \!cm$^{-1}$
(\ref{even}) and the other state is the $5p^5\,6f\,\,^2[3/2]_2$ state
with energy  $\rm{E} =94737.121 \ \rm{cm}^{-1}$ [Fig. \ref{fig:ExpApparatus}\,(a)]. \\
\indent From the energy shifts and splittings of the lines, one can deduce the scalar and tensor polarizabilites, and therefore identify respective contributions from nearby opposite-parity states \cite{CHLi,Rochester,BudkerDy}.\\ 
\subsection{Experimental Apparatus \& Methods} 
For the measurement of the Stark shift of the $6f$ states a  2+1 resonance-enhanced multi-photon ionization (REMPI) scheme using light at about 211\,nm was implemented\,\cite{Charalambidis}. A schematic diagram of the experimental setup is presented in Fig. \ref{fig:ExpApparatus}, and the details of the apparatus can be found in Refs. \cite{Samartzis1997,Kitsop2006,Kitsop2011}.\\ 
\indent An atomic beam of Xe atoms was supersonically expanded through a homemade piezoelectrically actuated nozzle valve ($\sim$1\,mm orifice diameter, backing pressure 1 atm.)\,\cite{Kitsop2011}. The atomic beam was skimmed and collimated, and $\sim$10\,cm from the nozzle was intersected at right angles by a linearly polarized pulsed-laser beam. In the interaction region, the atomic beam density was estimated to be 10$^{14}$\,atoms/cm$^3$, and the residual gas pressure in the vacuum chamber was $\sim10^{-6}$\,mbar. \\
\indent The 211\,nm light was generated with a tunable dye laser (Lambda Physik LPD300) that was pumped with a XeCl excimer laser LPX. The laser operated at a repetition rate of 10 Hz and had a pulsewidth of $\sim\!30$\,ns, while the laser linewidth was approximately 0.2\,cm$^{-1}$. To produce the desired fundamental wavelength ($\lambda$$\sim$\,422\,nm), Stilbene 3 dye was used. The typical output pulse energy was $\sim$7\,mJ. A BBO-II crystal was used to frequency double the output of the dye laser, producing thus the desired 211\,nm radiation. Typical pulse energies of the light at 211\,nm, were $\sim0.5$\,mJ. \\
\indent The particular states of interest were populated via two-photon absorption ($\lambda \sim 211$\,nm) starting from the ground state $5p^6$ ($^1S_0$). The laser/atomic beam interaction region lies in the focus of the ion-imaging system consisting of a single-electrode repeller grid (60\,mm outer diameter and 2\,mm aperture) and grounded extractor with a 20\,mm aperture (see Fig.~\ref{fig:ExpApparatus}\,b) \cite{Kitsop2006}. The 2\,mm hole on the repeller was also covered with a flat grid (1000\,lines-per-inch) to ensure homogeneity of the electric field. The distance between the repeller and the extractor was set at 2.5\,mm. High voltage, up to 9\,kV, is applied to the repeller grid using a high-voltage feed-through. Calibration of the supplied voltage was performed using a high-voltage probe (Tektronix P6015A). \\
\indent Following the two-photon excitation to a resonant state, the absorption of an additional 211\,nm photon ionizes the excited atom. The produced ions, located in the center of the repeller-extractor arrangement, are accelerated in a time-of-flight (TOF) apparatus toward a microchannel-plate (MCP) detector. In the presence of a dc static electric field (the one used for the acceleration in the TOF apparatus), the atomic levels are shifted due to the Stark effect, and therefore the frequency for the two-photon absorption changes. Maximizing the signal on the MCP detector by adjusting the wavelength of the laser, allows one to directly measure the resonant transition wavelength for a specific value of the electric field, and consequently measure the Stark effect of the state of interest. The MCP signal was displayed  on a digitizing oscilloscope (LeCroy WaveRunner 104MXi-A) which permitted signal averaging. Approximately 40 pulses were averaged per each data point. For each relevant transition, we performed three wavelength scans as a function of the electric field. At the end of each scan, the grating position of the dye laser for zero electric field was recorded. Determination of the wavelength of the transition in the absence of electric field was possible by adjusting the delay between the laser pulse and the rapid switching-on of the electric field after the excitation/ionization has taken place. The resolution on the position of the grating was 0.001\,nm.\\
\indent Finally, the orientation of the polarization of the laser beam with respect to the electric field was controlled with a $\lambda/2$ waveplate (Fig.\,\ref{fig:ExpApparatus}), allowing us to identify the observed shifted lines with particular Zeeman sublevels of the state under investigation, by exploiting the selection rules for transitions driven by two photons of the same color\,\cite{Bonin} for the specific geometry of our experiment (Table.\,\ref{t:selectionrules}). \\


\begin{table}[h!]
\caption{Selection rules for the $J=0\rightarrow J=2$ two-photon monochromatic excitation for the geometry shown in Fig.\,\ref{fig:ExpApparatus} (electric-field along $\hat{z}$-axis, and laser beam propagating along $\hat{y}$-axis). The quantization axis is chosen parallel to the electric-field axis ($z$-axis). Controlling the polarization of the laser beam, different Zeeman sublevels of the excited state with angular momentum $J=2$ can be identified.}
\label{t:selectionrules}
\begin{ruledtabular}
\begin{tabular}{ c | c c c}
Polarization & \multicolumn{3}{c}{Zeeman sublevels}\\
of laser beam & & &\\ [0.2ex]
\hline\\[-1.4ex]
$\hat{x}$   & $\Delta M=0$ &           & $\Delta M=\pm2$   \\[0.8ex]
$\hat{z}$  & $\Delta M=0$ &             &  \\[0.8ex]
$\frac{1}{\sqrt{2}}(\hat{x}+\hat{z})$  & $\Delta M=0$ &$\Delta M=\pm1$ & $\Delta M=\pm2$ \\[0.8ex]
\end{tabular}
\end{ruledtabular}
\end{table}
\subsection{Electric-Field Calibration: Stark Effect of the $n=2$ states in hydrogen} 
\label{seq:EfieldCalib}
\indent To verify the performance of the apparatus for the investigation of the Stark effect in the excited states of Xe we measured the Stark effect on the two-photon $1s-2s$ transition of the hydrogen atom (at 243\,nm). The Stark splitting and mixing of the 2s$_{1/2}$ and 2p$_{\{1/2,3/2\}}$ states of atomic H can be calculated exactly and therefore provide an accurate determination of the electric field strength in the interaction region \cite{Luders,Kollath,DeLaRosa}.\\
\indent \emph{Method -} The presence of diffusion-pump oil in the chamber is the main source of H atoms in our experiment. A (2+1) REMPI scheme at 243\,nm was used to excite and ionize H atoms produced by the photodissociation of pump-oil hydrocarbons. The photodissociation produces a top-hat Doppler frequency profile approximately $4$\,cm$^{-1}$ wide. For low electric fields, the large Doppler widths on the observed line mask the expected Stark shifts of the $n=2$ states (of the order of a few wavenumbers). For that reason, a Doppler-free (2+1) REMPI scheme was adopted to produce and ionize H atoms at a single laser wavelength regardless of the atoms' lab-frame velocity (Doppler shift) using circularly polarized light\,\cite{Grynberg,ZareDF}. Doppler-free excitation produces narrow absorption features, increasing thus the electric-field detection sensitivity through the Stark effect \cite{DeLaRosa}. By varying the intensity of the laser we have verified that light-induced (ac) Stark shifts are negligible. \\
\indent  For the production of 243\,nm radiation, Coumarin 120 dye was used, yielding energies of $\sim1$\,mJ per pulse at 243\,nm (after the frequency-doubling stage). For the realization of a Doppler-free REMPI scheme, additional optics were used (see, Fig.\,\ref{fig:ExpApparatus}\,(b)). The linear polarized beam was converted to a circularly polarized beam using a $\lambda/4$ waveplate, and focused in the center of the repeller/extractor geometry using a $f=15$\,cm lens. Upon exiting the chamber, the beam passed through another lens ($f=20$\,cm), creating a collimated beam, which was then reflected back to the chamber with the use of a plane mirror (dashed box in Fig. \ref{fig:ExpApparatus}\,(b)). Between the lens and the retroreflecting mirror, an additional $\lambda/4$ waveplate was used to control the polarization of the reflected light. Spatial overlap in the focus was ensured by directing the reflected beam back through the initial beam path, and by adjusting the position of the retroreflecting-mirror to produce a beam matched with the input one. Creating counter-propagating beams by using a mirror reflection, introduces a delay between the two beams (in our case $\sim2$\,ns) but temporal overlap was ensured by the fact that the laser pulse width is $\sim$30\,ns, maintaining, thus, a high ionization efficiency \cite{ZareDF}. Calibration of the polarization of the counterpropagating beams was performed by exploiting the selection rules for two-photon transitions. The 243\,nm resonant signal disappears when the counter propagating beams have opposite helicities.\\
\begin{figure}[h!]
\centering
\epsfxsize=3.4in
\epsffile{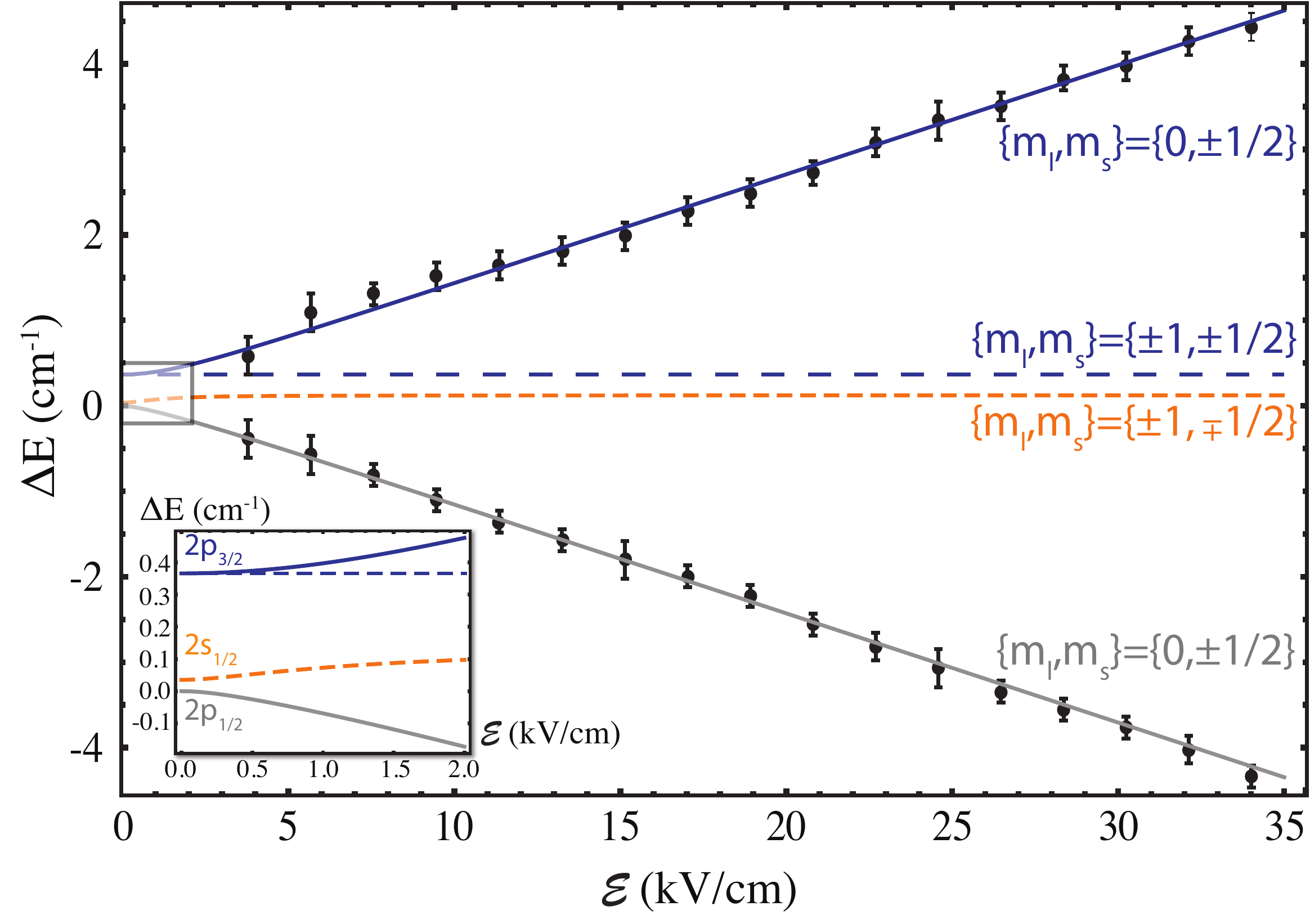}
\caption{\small{(color online). Theory and experimental data for the Stark effect of the two-photon $1s-2s$ transition of hydrogen. The calculated shifts for the mixed $2s_{1/2}$ (orange, dashed), $2p_{1/2}$ (gray, solid), $2p_{3/2}$ (blue, solid) states are presented as a function of the electric field strength. The zero value in the energy axis which corresponds to the unperturbed energy of the $2p_{1/2}$ state, and consequently the energy splittings between the $2p_{1/2}$ state and the $2s_{1/2}$ and $2p_{3/2}$ states represent the Lamb shift and the fine-structure splitting, respectively\,\cite{NIST} (see inset). The unperturbed $2p_{3/2},m_j=\pm3/2$ (blue, dashed) state is also presented. The hyperfine structure is neglected under our experimental conditions; see text for discussion.}}
\label{fig:StarkHydrogen}
\end{figure}
\indent \emph{Results \& Analysis -} In Fig. \ref{fig:StarkHydrogen} we present the results of the calibration experiment. The solid lines (Fig. \ref{fig:StarkHydrogen}) represent the theoretical calculations for the Stark shift of the $n=2$ states in H as a function of the electric field. \\ 
\indent In the absence of electric field, only the two-photon transition to the $2s_{1/2}$ state is allowed. With increasing electric field, the $2p_{1/2}$ component of the mixed state appears and is red-shifted, while the $2p_{3/2}$ component of the mixed state is blue-shifted. For sufficiently high electric fields ($\geqslant4$\,kV/cm) the fine-structure is negligible and $j$ is not a good quantum number. In two-photon processes, for collinear and of equal-frequency photons, the selection rules state that the photons cannot participate in any process that would require them to be in a state of total angular momentum one\,\cite{Bonin}. Therefore, as a function of the electric field, the probability of exciting into states characterized by angular momentum one should tend to zero. That is apparent in Fig.\,1(b) of Ref.\,\cite{DeLaRosa}, where the normalized two-photon absorption line-intensities are presented. Note that for fields higher than 5\,kV/cm the $2s_{1/2}$ intensity drops to less than $\sim$5\% of its initial value. This fact explains the inability to observe transitions to the $2s_{1/2}$ in our experiment, where for low electric fields the signal-to-noise ratio was approximately SNR$\sim$5. For this work, we chose to take data at electric fields higher than 4\,kV/cm for better resolution. Moreover, the interaction of the atom with the electric field does not mix states of different $m_j$ when the quantization axis is chosen parallel to the electric field\,\cite{note1}, and therefore the $2p_{3/2}$,$m_{j=\pm3/2}$ sublevels remain unperturbed by the presence of the electric field (as seen in Fig.\,\ref{fig:StarkHydrogen}). In addition, note that for the polarizations chosen in our experiment, the transition to the $2p_{3/2}$,$m_{j=\pm3/2}$ could not be observed. Furthermore, the hyperfine structure of hydrogen is neglected, as the hyperfine splittings of the $n=2$ levels of hydrogen are much smaller than our experimental energy resolution\,\cite{Bethe}, while Stark-shifts of the hyperfine structure for the electric-field strengths used, are negligible\,\cite{Anderson}.\\
\indent The dye laser's grating positions corresponding to signal peaks were recorded for each value of the electric field. Each data point shown in Fig. \ref{fig:StarkHydrogen} represents the average from three consecutive scans. The error bar for each point shown in Fig.\,\ref{fig:StarkHydrogen} includes the correlated systematics which correspond to the uncertainty in the reading of the grating position. \\
\indent The experimental data are in good agreement with the theoretical predictions, as seen in Fig.\,\ref{fig:StarkHydrogen}, verifying the performance of the apparatus for the Stark-shift measurements of the $6f$ states of Xe presented in the next section.  
\section{Results and Discussion}
\label{sec:Results}
\subsection{Stark Effect of $6f$ states} 
\begin{figure}[h!]
\begin{center}
$\begin{array}{c@{\hspace{0.3in}}c}
\multicolumn{1}{l}{\mbox{\bf (a)}} &\\
\epsfxsize=3.3in
\epsffile{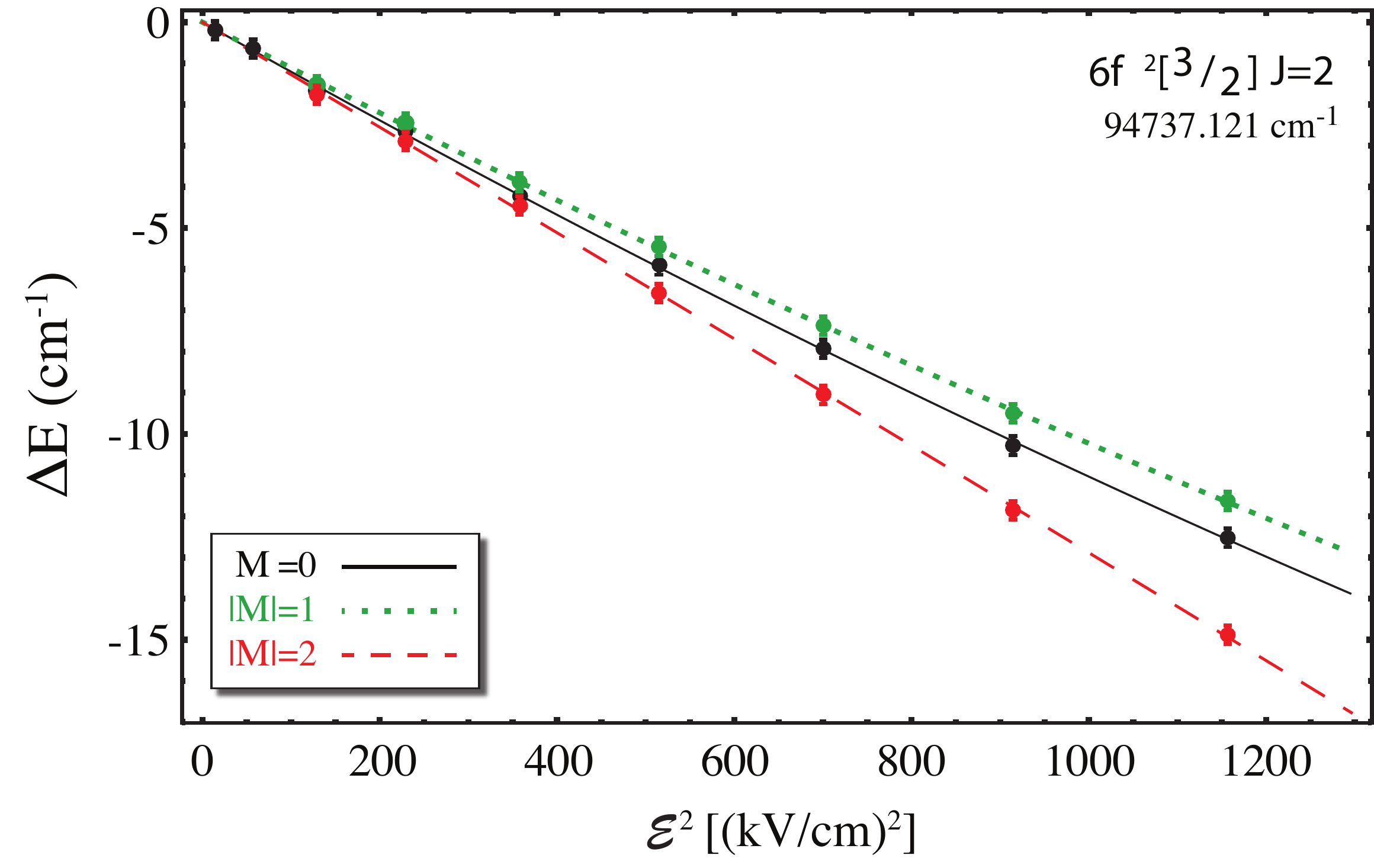} &\\
	\multicolumn{1}{l}{\mbox{\bf (b)}} &\\ 
	\epsfxsize=3.3in
	\epsffile{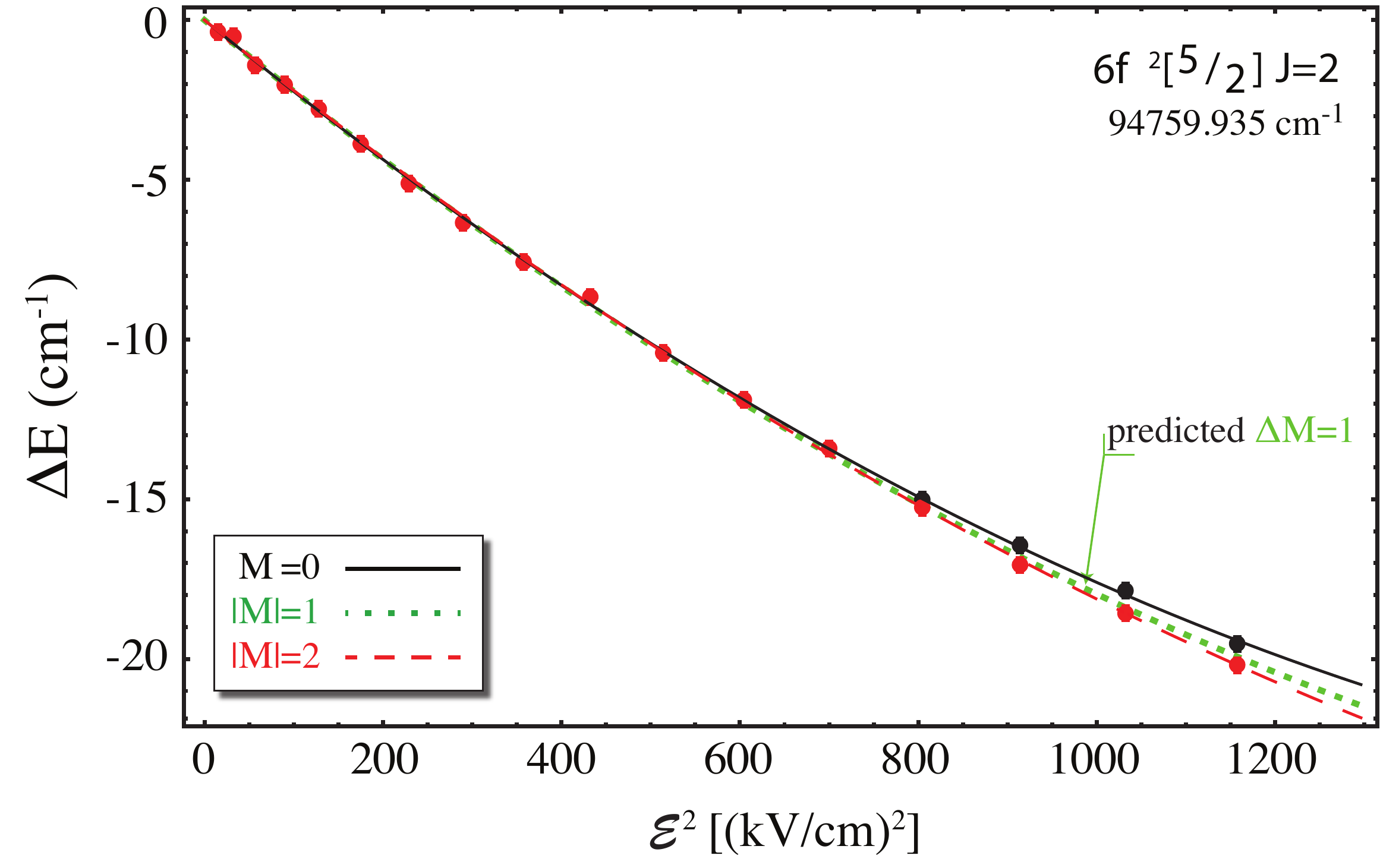} \\ 
\end{array}$
\end{center}
\caption{\small{(color online) Experimental data of the Stark-shift of states (a) $6f$ $^2[3/2]$ $J=2$ $(94737.121$ \!cm$^{-1})$; $M=0$ (solid line), $|M|=1$ (dotted line), $|M|=2$ (dashed line) Zeeman sublevels and (b) $6f$ $^2[5/2]$ $J=2$ $(94759.935$ \!cm$^{-1})$, $M=0$ (solid line), $|M|=2$ (dashed line) Zeeman sublevels, along with the theoretical prediction for the Stark-shift of the $|M|=1$ Zeeman sublevels (dotted line) using the polarizability and hyperpolarizability obtained from fitting Eq.\,\eqref{eq:StarkEffectEq} to the observed energy-shifts of the $M=0$ and $|M|=2$ Zeeman sublevels. In both cases, the different Zeeman sublevels were identified by exploiting the degenerate two-photon selection rules (Table\,\ref{t:selectionrules}), made possible via the control of the polarization state of the laser beam.}}
\label{fig:Stark6f}
\end{figure}
\indent In Fig.\,\ref{fig:Stark6f} we present the Stark-shift measurements of the $6f$ states. There are two important features of these shifts which should be noted. First, in the case of the $6f$ $^2[5/2]_2$ state (\ref{even}), the magnitude of the shift (15-20 cm$^{-1}$) is much larger than the energy interval between states (\ref{odd}) and
(\ref{even}). Therefore, the Stark-shift should be generally treated non-perturbatively via
matrix diagonalization. Second, the shifts for both states are negative, i.e., the energy intervals
between the ground state and the excited $6f$ states decrease with increasing
electric field. This means that the shifts are dominated by mixing with odd-parity
states above the $6f$ states. The largest contribution should come from the
$6g$ states. Their position is not known experimentally but it is
natural to expect that the energies of the $6g$ states are close
to the energies of the $6f$ states. Indeed, both configurations correspond to almost
pure Coulomb states screened from the atomic core by a centrifugal
barrier. Their energies ($E \approx -Ry/n^2$, where $Ry$ is the Rydberg constant and $n$ is the principal quantum number) in the Coulomb limit do not
depend on angular momentum, as confirmed by Hartree-Fock
calculations. \\
\indent The energy interval between the $6f$ and $6g$ states is not known, and for that reason it is treated as a fitting parameter in our analysis of the experimental data. Using the Stark-shift measurements for the $6f$ states, a range of values for the energy interval between the $6f$ and $6g$ states can be obtained. In particular we use the Stark-shift matrices presented in Tables\,\ref{t:m3} and\,\ref{t:m2}, and, for example, focus our analysis on the Stark-shift data for the $|M|=2$ Zeeman sublevels for both $6f$ states (Fig.\,\ref{fig:Stark6f}). \\
\indent The electric-dipole matrix element $d_1$ appearing in Table\,\ref{t:m3}, is given by Eq.\,\eqref{eq:e1z}. The electric-dipole matrix elements $d_2$ and $d$, appearing in Tables\,\ref{t:m3} and\,\ref{t:m2}, include all possible transitions from the $6f\,^2[5/2]_{2}$ and the $6f\,^2[3/2]_{2}$ states to the $6g$ states, respectively\,\cite{note2}. Theoretical estimates for these electric-dipole transition amplitudes, $d_2$ and $d$, suggest that $d_2\approx d\approx20(5)\,ea_0\rightarrow0.86(22)\,\rm{cm}^{-1}/(\rm{kV}/\rm{cm})$. Therefore, the small value of the electric-dipole transition amplitude between the $6f$ and $10s$ states, $d1$ [Eq.\eqref{eq:e1z}], should have little effect on the analysis of the measurements. This is verified by setting $d_1=0$, where we see that the Stark-shift analysis of the measurements on the $6f\,^2[5/2]_{2}$ [Fig.\,\ref{fig:Stark6f}\,(a)] is not significantly affected. Moreover, we neglect the contribution of the $10s$ state in the Stark-shift of the $6f$\,$^2[3/2]_{2}$ state (see Table\,\ref{t:m2}). This state is not so close to the $10s$ state, and thus, the corresponding mixing is further suppressed by orders of magnitude. \\
\begin{table}[ht!]
\caption{Stark shift matrix for the $6f$\,$^2[5/2]_{J=2}$ state with $E=94759.935 \
  \rm{cm}^{-1}$. $d_1$ and $d_2$ are the electric dipole transition
  amplitudes to the $10s$ and $6g$ states (see text); $k_f$ and $k_g$
  are the second-order energy corrections due to transitions to the states
  not included in the matrix; $\delta$ is the $6f$ - $6g$ energy
  interval; $\mathcal{E}$ is the static electric field strength. Transition
  amplitudes $d_1$ and $d_2$ are taken from
  calculations; $k_f$, $k_g$ and $\delta$ are treated as fitting parameters.}
\label{t:m3}
\begin{ruledtabular}
\begin{tabular}{r|lll}
States & \multicolumn{1}{c}{$6f$} & \multicolumn{1}{c}{$10s$} &
\multicolumn{1}{c}{$6g$} \\
\hline
$6f$   & $k_f\mathcal{E}^2$ & $d_1\mathcal{E}$   & $d_2\mathcal{E}$   \\
$10s$  & $d_1\mathcal{E}$   & 0        &  0 \\
$6g$   & $d_2\mathcal{E}$   & 0        & $\delta + k_g\mathcal{E}^2$ \\
\end{tabular}
\end{ruledtabular}
\end{table}
\begin{table}[ht!]
\caption{Stark-shift matrix for the $6f$\,$^2[3/2]_{J=2}$  state with $E=94737.121 \
  \rm{cm}^{-1}$. $d$ is the electric dipole transition
  amplitude to the $6g$ states (see text); $k_f$ and $k_g$
  are second-order energy corrections due to transitions to the states
  not included in the matrix; $\delta$ is the $6f$ - $6g$ energy
  interval and $\mathcal{E}$ is the static electric field strength. Transition amplitude $d$ is taken from
  calculations; $k_f$, $k_g$ and $\delta$ are treated as fitting parameters.}
\label{t:m2}
\begin{ruledtabular}
\begin{tabular}{r|ll}
States & \multicolumn{1}{c}{$6f$} &
\multicolumn{1}{c}{$6g$} \\
\hline
$6f$   & $k_f\mathcal{E}^2$ & $d\mathcal{E}$   \\
$6g$   & $d\mathcal{E}$   & $\delta$ + $k_g\mathcal{E}^2$ \\
\end{tabular}
\end{ruledtabular}
\end{table}

\indent We proceed with the analysis of the Stark-shift measurements for the $|M|=2$ Zeeman subleveles of both $6f$ states by imposing two extra conditions. The parameters $k_f$ and $k_g$, appearing in the Stark-shift analysis for both $6f$ states (Tables\,\ref{t:m3} and\,\ref{t:m2}) are second-order energy corrections, due to transitions to states not included in the matrices. In particular, the $k_g$ parameter appearing in both matrices, is the same as it corresponds to the same set of the $6g$ states. The $k_f$ and $k_g$ parameters model the quadratic behavior of the Stark-shift measurements while the other terms describe the deviation from the quadratic behavior. The parameter $\delta$, which corresponds to the energy interval between the $6g$ state and the $6f\,^2[5/2]_{2}$ state ($\Delta \delta =23 \,\rm{cm}^{-1}$ is the experimental value for the energy interval between the two $6f$ states), is included in both Stark-shift-matrix diagonalizations. For Stark shifts smaller than $\delta$, one expects a quadratic Stark effect, while for larger values, a linear behavior. Therefore, the parameter $\delta$ is constrained by the deviation from the quadratic behavior, and thus, its value is directly linked to the value of $k_f$. Furthermore, one sees from the analysis that for large values of $d_2$ and $d$, one gets $k_f>0$ as expected, while for small values, one gets $k_f<0$. Imposing these constraints on the analysis of the Stark-shift data for both $6f$ states, along with the theoretical estimations for the $d_2\approx d\approx20(5)\,ea_0$, for which $k_f>0$, we find an energy interval $\delta=30(10)$\,cm$^{-1}$. \\
\indent For the physically expected range of values for $k_f$ and $\delta$, we see from the theoretical analysis that $k_g$ is negative. This is probably due to a contribution from the $6h$ states, similar to the negative Stark shift for the $6f$ states arising due to the contributions of the $6g$ states. Therefore, the measured Stark shift of the two $6f$ states reveals the presence of the $6g$ and $6h$ states. \\
\indent Finally, the data on the Stark-shift for the state (\ref{even}) can be used to
put a limit on the weak mixing of the degenerate states (\ref{odd}) and
(\ref{even}). This is because the weak matrix element between these states
and the amplitude of the electric dipole transition between them, are
affected by configuration mixing in a similar way. Indeed, if
both matrix elements are dominated by the $10s$ - $np$ transitions, then
the value of both matrix elements is determined by the admixture of
the $5p^5np$ configurations to the $5p^56f$ configuration. This is estimated by the value of the $d_1$ parameter in the Stark shift matrix (Table\,\ref{t:m3}). If the calculated value of $d_1$\,($\approx0.5\,ea_0$) is used, very good fitting of the experimental Stark-shift data can be achieved. However, since this matrix element is not zero due to configuration mixing only, the uncertainty in its value is large. The same is true for the weak matrix element $W$ [Eq.\,\eqref{eq:res}]. As a way to estimate this uncertainty we check how much one can change the value of $d_1$ without affecting the fitting of the Stark-shift data. It turns out that the analysis becomes unstable for $d_1 >1\,ea_0$. The instability manifests itself in the need to use large unrealistic values for the fitting parameters, and deteriorating of the quality of fitting. Taking the value $d_1=1\,ea_0$ as the upper limit for the electric dipole transition amplitude, we note that it is about 14 times smaller than the $10s$ - $10p_{1/2}$ electric dipole amplitude. Assuming the same ratio for the weak matrix elements and using:
\begin{equation}
\langle 10s |H_W| 10p_{1/2} \rangle = 0.12 \times 10^{-13} (-Q_W/N)
i\frac{e^2}{a_0},
\label{eq:ps}
\end{equation}
obtained in the RPA calculations, we get for the weak matrix element
between the degenerate states (\ref{odd}) and (\ref{even}): $|W| < 0.8\times 10^{-15} (-Q_W/N) \ ie^2/a_0\equiv 5$\,Hz in agreement with the calculated
value [Eq.\,\eqref{eq:res}]. \\
\subsection{Polarizabilities and hyperpolarizabilities of the 6$f$ states}
\label{subsec:Polarizab}
\indent In the case of barium (Ba)\,\cite{CHLi} and samarium (Sm)\,\cite{Rochester}, Stark-shift measurements of various states were used to estimate the value of the reduced matrix-elements of the dipole operator to near opposite-parity ``partner" states. This proved to be possible, in most cases, because the observed shifts and sublevel splittings for each state under investigation, were the result of contribution by one or a few dominating near opposite-parity partner-states in each case (which dominate due to the smallness of the energy-difference denominator). In the case of Xe it appears that this is not possible, and a more refined level of theoretical calculations is required to determine all the reduced matrix elements between the $6f$ states and the $6g$ states. \\
\indent In this section, we present an analysis of the Stark-shifts for all the observed Zeeman sublevels of the $6f$ states, presented in Fig.\,\ref{fig:Stark6f}, in terms of the polarizabilities and hyperpolarizabilities, following closely the work described in Ref.\,\cite{CHLi} (and references therein).\\
\indent The polarizability, $\alpha_{\eta J M}$, and hyperpolarizability, $\gamma_{\eta J M}$, of the state $|\eta J M\rangle$ is determined from the dependence of the observed energy shifts on the electric field by fitting the data with the following equation:
\begin{equation}
\Delta\text{E}_{\eta J M}=-\frac{1}{2}\alpha_{\eta J M} E^2-\frac{1}{4!}\gamma_{\eta J M} E^4.
\label{eq:StarkEffectEq}
\end{equation}
The polarizability, $\alpha_{\eta J M}$, can be expressed in terms of its scalar and tensor parts as:
\begin{equation}\label{eq:alphaPol}
\alpha_{\eta J M}=\alpha_{0,\eta J M}+\alpha_{2,\eta J M}\frac{3M^2-J(J+1)}{J(2J-1)},
\end{equation}
and the hyperpolarizability, $\gamma_{\eta J M}$, can be expressed in terms of its scalar and tensor parts as:
\begin{widetext}
\begin{align}\label{eq:gammaPol}
\gamma_{\eta J M}=& \, \gamma_{0,\eta J M}+\gamma_{2,\eta J M}\frac{3M^2-J(J+1)}{J(2J-1)} \nonumber \\
& +\gamma_{4,\eta J M}\frac{35 M^4+[25-30 J (J+1)] M^2+J(J-1)(J+1) (J+2)}{J (2 J-1)(2 J-3) (2 J-2)}.
\end{align}
\end{widetext}

\indent In Tables\,\ref{Polariz} and \ref{HyperPolariz} we present the polarizabilities and hyperpolarizabilities of the 6$f$ states obtained in this work. Note that the polarizability of the ground state of Xe is neglected, as the energy shifts of the $5p^6$ ground state for the maximum electric-fields used in our studies are of the order of a few $\sim$MHz\,\cite{Nicklass} and therefore are negligible compared to the observed shifts.\\
\indent Using the values for the $\alpha_0$ and $\alpha_2$ of the $6f$ $^2[5/2]$ $J=2$ state (Table\,\ref{Polariz}), we predict the energy-shift for the $|M|=1$ Zeeman sublevels that we were not able to identify by exploiting the selection rules of the two-photon transition (see Table\,\ref{t:selectionrules}). In addition, because a large deviation from quadratic behavior is observed in the energy-shift of the $6f$ $^2[5/2]_2$, we use the hyperpolarizabilities $\gamma$ obtained from the energy-shifts of the $|M|=0,2$ sublevels (Table\,\ref{HyperPolariz}) to estimate $\gamma(|M|=1)$. In Fig.\,\ref{fig:Stark6f}\,(b) the prediction for the energy shift of the $|M|=1$ sublevel is presented. The predicted overlap between the $|M|=1$ and the $|M|=2$ Zeeman sublevels explains the inability to identify the Stark-shift of the $|M|=1$ sublevels considering the experimental resolution of our apparatus (note that the uncertainty in the reading of the dye laser's grating position was $\sim0.22$\,cm$^{-1}$). \\
\indent The Stark-shifts of the observed Zeeman sublevels of the $6f[3/2]_2$ state, cannot be explained using a similar procedure as the one followed for analyzing the Stark-shift data of the $6f[3/2]_2$ state. The tensor energy-shifts are expected to be linear in $M^2$, where the observed data show a different behavior. Therefore, different coupled partner-states for each $|M|$ sublevels are required to explain the observed spectrum. We analyze separately each observed sublevel of the state $6f[3/2]_2$ using Eqs.\,\eqref{eq:StarkEffectEq}, to obtain their polarizabilities and hyperpolarizabilities. The results are presented in Tables\,\ref{Polariz} and \ref{HyperPolariz}.\\
\indent Finally, note that the polarizabilities presented in Tables\,\ref{Polariz} and \ref{HyperPolariz} are among the largest measured for atomic systems in states with principal quantum number $n<10$ \cite{CHLi}.\\

\begin{table}[h!]
\begin{center}
\caption{Observed polarizabilities for each Zeeman sublevel of the 6f[5/2]$_2$ states, in units of MHz/(kV/cm)$^2$.} 
\renewcommand{\arraystretch}{1.6}
  \begin{tabular}{  c || c | c | c}
    \hline\hline
   \textbf{State} & $\alpha(M=0)$ & $\alpha(|M|=1)$ &$\alpha(|M|=2)$ \\ [5pt] \hline\hline
    6f[3/2]$_2$ \quad    &  728(9)	& 671(8)	 &762(11) 	 \\ [5pt]
    6f[5/2]$_2$ \quad    & 1370(11) 	& -	 &	1344(12)	\\    \hline \hline
          \end{tabular}
              \label{Polariz}
              \end{center}
\end{table}
\begin{table}[h!]
\begin{center}
\caption{Observed hyperpolarizabilities for each Zeeman sublevel of the 6$f$ states, in units of kHz/(kV/cm)$^4$.} 
\renewcommand{\arraystretch}{1.6}
  \begin{tabular}{  c || c | c  | c}
    \hline\hline
               \textbf{State}    & $\gamma(M=0)$& $\gamma(|M|=1)$ & $\gamma(|M|=2)$ \\[5pt] \hline\hline
    6f[3/2]$_2$ \quad    & -787(107) &  -689(95)  		&125(131) \\ [5pt]
     6f[5/2]$_2$ \quad     & -3776(140) & - 				&-3095(163)  \\[5pt]               
    \hline\hline
          \end{tabular}
    \label{HyperPolariz}
    \end{center}
\end{table}

\section{Conclusions}
We have studied the properties of a pair of nearly degenerate opposite-parity states in atomic Xe, namely the $5p^5 10s \, ^2[3/2]_2^o$ (\ref{odd}) and $5p^5  6f \, ^2[5/2]_2$ (\ref{even}) states, which are of interest for P- and P,T-odd experiments. Theoretical calculations of the width of the states, and the value for the weak matrix element between these states are presented. Furthermore, we studied the Stark-shift of the even-parity $6f$ state and put a limit on the weak mixing of the (\ref{odd}) and (\ref{even}) near-degenerate states, of $|W|<5$\,Hz. The analysis of the experimental Stark-shift measurements also revealed the presence of the $6g$ and $6h$ states in atomic Xe, which have been unobserved so far.\\
\acknowledgements
The experimental work was conducted at the Ultraviolet Laser Facility at
FORTH-IESL, supported in part by the EC FP7 project LASERLAB-EUROPE (grant
agreement no. 284464). This work is supported by the European Research Council (ERC) grant
TRICEPS (GA No. 207542), by the National Strategic
Reference Framework (NSRF) grant Heracleitus II
(MIS 349309-PE1.30) co-financed by EU (European
Social Fund) and Greek national funds, by the Australian Research Council, by the Marie Curie Initial Training network ICONIC
(PITN-GA-2009-238671) and by the Greek Secretariat of Research and Technology
(GSRT) under the THALIS program ISEPUMA. PCS gratefully acknowledges
financial support from a European Union Marie Curie Reintegration Grant (``GPSDI", Grant
No. PIRG07-GA-2010-268305). TNK also acknowledges support by the Alexander
von Humboldt foundation. JS was
supported in part by NSF grant PHY1068065. DB was supported by NSF, grant PHY-1068875.



\begin{thebibliography}{99}
\bibitem{note1} The interaction of the atom with a uniform dc electric field along the $z$-axis is described by the Hamiltonian $\mathcal{H}_{\mathcal{E}}=-\mathcal{E}D_z$.  $\mathcal{H}_{\mathcal{E}}$ is invariant under rotations about the $z$-axis, and therefore only $\Delta M=0$ transitions are allowed, and there is no $p-p$ mixing by the electric field.
\bibitem{note2} In the $jl$ coupling scheme there is one $6g$ state with total angular momentum $J=2$ and three $6g$ states with $J=3$ (see Ref. \cite{Sob}).
\end{thebibliography}
\end{document}